# Uniaxial negative thermal expansion in a weak-itinerant-ferromagnetic phase of CoZr$_2$H$_{3.49}$

Yuto Watanabe[1*], Kota Suzuki[2], Takayoshi Katase[2,3], Akira Miura[4], Aichi Yamashita[1] and Yoshikazu Mizuguchi[1*]

[1] Department of Physics, Tokyo Metropolitan University, 1-1 Minami-Osawa, Hachioji, Tokyo 192-0397, Japan

[2] MDX Research Center for Element Strategy, Institute of Integrated Research, Institute of Science Tokyo, 4259 Nagatsuta, Midori-ku, Yokohama, Kanagawa 226-8501, Japan

[3] Materials and Structures Laboratory, Institute of Integrated Research, Institute of Science Tokyo, 4259 Nagatsuta, Midori-ku, Yokohama, Kanagawa 226-8501, Japan

[4] Faculty of Engineering, Hokkaido University, Kita13, Nishi8, kita-ku, Sapporo, Hokkaido 060-8628, Japan

**ABSTRACT:** We discovered unique uniaxial negative thermal expansion (NTE) behavior for a weak-itinerant-ferromagnetic phase of CoZr$_2$H$_{3.49}$. CoZr$_2$ is known as a superconductor exhibiting uniaxial NTE along the *c*-axis, which is called anomalous thermal expansion (ATE). Additionally, CoZr$_2$ is also known as a well-absorbent of hydrogen, and hydrogen insertion raises weak-itinerant ferromagnetism instead of superconductivity. However, the influence of hydrogen insertion on ATE behavior in this system is still unclear. To investigate it, we performed powder synchrotron X-ray diffraction (SXRD) for CoZr$_2$H$_{3.49}$. Through Arrott plots analysis, we determined the Curie temperature ($T_C$) to be 139 K, and the Rhodes–Wohlfarth ratio was estimated to be 3.49, which clearly exceeds 1, suggesting the itinerancy of emerging ferromagnetism. Temperature dependencies of lattice constants *a* and *c* were extracted from powder SXRD analyses, and we revealed that lattice constant *c* exhibited NTE behavior below $T_C$. The uniaxial NTE behavior along the *c*-axis can be understood by sharpening an antibonding Co3$dz^2$ partial density of states near the Fermi level, linked to the expansion of a one-dimensional Co–Co chain running parallel to the *c*-axis.

## 1. INTRODUCTION

Studies on negative thermal expansion (NTE) materials, which shrink on heating, have undergone dramatic development over the past decade due to strong demand from contemporary industrial applications.[1,2] Controlling the coefficient of thermal expansion is a key technology to prevent fatal errors caused by linear distortion.[3] NTE materials possess significant potential to control them; therefore, research on NTE materials has garnered attention.[4–6] Since the discovery of NTE behavior in Zr/HfW$_2$O$_8$ with a wide NTE temperature range of 0.3–1050 K in 1996,[7] various NTE materials have been rapidly developed. Mechanisms of NTE behavior can be classified into two types: a conventional type, driven by phonons with a flexible-framework structure, and a phase-transition type, driven by various phase transitions, including structural, magnetic, ferroelectric, charge-transfer, and metal-insulator transitions.[8] The operating temperature range of NTE behavior is different between the two types. For the conventional type, the NTE behavior can be commonly observed in a very wide temperature range. In contrast, for the phase-transition type, NTE behavior can only be observed below the phase transition temperature.

As one of the phase-transition types of NTE mechanism, superconducting transitions are known as a driving force for spontaneous strain.[4] For instance, NdFeAsO$_{0.89}$F$_{0.11}$[9] and La$_{1.85}$Sr$_{0.15}$CuO$_4$[10] exhibit NTE behavior below the superconducting transition temperature ($T_{SC}$). NTE behavior coupled with the superconducting transition usually occurs below $T_{SC}$. In 2022, however, the CoZr$_2$ superconductor with $T_{SC}$ = 6 K was found to exhibit uniaxial NTE behavior.[11] The crystal structure belongs to the CuAl$_2$-type tetragonal space group $I4/mcm$ (no. 140), and the NTE behavior appears only along the *c*-axis, whereas the *a*-axis exhibits positive thermal expansion (PTE). Because of the anisotropy of thermal expansion properties, the uniaxial NTE behavior is called anomalous thermal expansion (ATE). The discovery of ATE behavior in the CoZr$_2$ superconductor has led to further exploration of other *Tr*Zr$_2$ compounds (*Tr*: transition metal) with an abundant variation of *Tr* elements.[12–15] Notably, these studies revealed an empirical tendency that a large $c/a$ ratio ($c/a > \sim0.84$) is advantageous for the emergence of ATE.[13] Additionally, for superconductivity, the lattice constant *c* and $T_{SC}$ have a positive correlation.[16] Another possible way to change the ATE and superconducting properties is by inserting hydrogen because zirconium-based compounds are known to be well-absorbent of hydrogen.[17–21] Hydrogen insertion has a significant potential in changing the characteristics of materials, for instance, electric, magnetic, and structural properties.[22,23] Relevant to the *Tr*Zr$_2$ superconductors, a weak-itinerant ferromagnetism instead of superconductivity was discovered in a hydrogen-inserted compound CoZr$_2$H$_{4.8}$ with $T_C$ = 128 K ($T_C$: Curie temperature),[24] following theoretical expectation.[25] The drastic physical property changes by inserting hydrogen suggest that the ATE behavior also should be influenced because the mechanisms of NTE correlate with electronic and/or phononic properties.

Therefore, we have investigated the thermal expansion behavior of hydrogen-inserted CoZr$_2$H$_x$ for *x* = 3.49. Although the hydrogen concentration of the obtained sample is a little lower than that of the sample reported in Ref. 24 (*x* = 4.8, $T_C$ = 128 K), we confirmed that the synthesized CoZr$_2$H$_{3.49}$ also exhibits a weak-itinerant-ferromagnetic transition with a higher $T_C$ of 139 K. We observed uniaxial NTE behavior along the *c*-axis only in its weak-itinerant-ferromagnetic phase. In this article, we report on the observation of the ATE behavior in CoZr$_2$H$_{3.49}$ and propose a tentative scenario to explain the ATE behavior in conjunction with the weak-itinerant-ferromagnetic transition.

## 2. EXPERIMENTAL METHODS

### 2.1 Synthesis and hydrogenation.

A polycrystalline sample of $CoZr_2$ was synthesized using an electric arc furnace. Starting materials of Co wires (99.995%, Nilaco) and Zr plates (99.2%, Nilaco) were put on a water-cooled Cu hearth, and they were in an Ar atmosphere during the melting procedure. Before starting the melting process, a Ti ball was heated to absorb any residual oxygen. After that, Co wires and Zr plates were melted. The ball-shaped sample was melted several times to improve its homogeneity. Synthesized $CoZr_2$ was ground to make it a fine powder, and part of it was treated by hydrogenation. The hydrogenation was performed by heating the powder sample covered by Mo foil (99.95%, Nilaco) at 400°C under 0.5 MPa of hydrogen gas for 1 hour. The hydrogen-inserted powder sample of $CoZr_2H_x$ and the rest of the non-hydrogenated $CoZr_2$ were sintered under high pressure using a cubic-anvil type 180-ton press (C&T Factory). Pelletized samples with a diameter of 5 mm were encapsulated into a high-pressure cell, which consists of a BN sample capsule, a carbon heater capsule, electrodes, and a pyrophyllite cubic cell. The pelletized sample covered by the cubic cell was heated at 300°C under 1.5 GPa for 20 minutes. We performed the following characterization using the high-pressure-annealed sample $CoZr_2$ and $CoZr_2H_x$.

### 2.2 Characterization.

Crystal structures of the synthesized materials were determined by powder synchrotron X-ray diffraction (SXRD) with a wavelength of 0.354443 Å performed at BL13XU in SPring-8 (Proposal no.:2025A0214). Powder samples were sealed into an evacuated quartz capillary with a diameter of 0.2 mm. Sample temperature was controlled by blowing nitrogen gas through a temperature-controlled gas-handling system with a speed of 80 K/min in a range of 100–300 K. Collected powder SXRD patterns were refined by the Rietveld method using the RIETAN-FP,[26] and refined crystal structures were depicted by VESTA3.[27] Actual chemical compositions of Co and Zr were determined by energy-dispersive X-ray spectroscopy (EDX) on a scanning electron microscope TM-3030plus (Hitachi High-Tech) equipped with computer software SwiftED (Oxford). The actual hydrogen concentration was estimated by thermal desorption spectroscopy (TDS). We confirmed that the actual hydrogen concentration of the obtained sample was $x$ = 3.49, and the actual composition ratios of Co:Zr for $CoZr_2$ and $CoZr_2H_{3.49}$ agree with the nominal composition ratio. Temperature dependencies of electrical resistivity $\rho(T)$ were measured using a PPMS Dynacool (Quantum Design) system by the conventional four-probe method with an alternating current of 2 mA. Ag paste and Au wires were used to connect the samples with a sample puck. Temperature dependencies of magnetism $M(T)$ were measured using MPMS3 (Quantum Design) in vibrating sample magnetometer mode. An effective magnetic field ($H_{eff}$), which considers a demagnetizing effect, can be expressed as the following equation:

$$H_{\text{eff}} = H - 4\pi MN, \quad (1)$$

where $H$ and $N$ are applied magnetic field and demagnetizing factor, respectively. The factor $N$ depends on the sample geometry and the direction of $H$. In a case of cuboid-shaped samples, factor $N$ is calculated by the following equation:[28]

$$N = \frac{4lw}{4lw + 3wt + 3lt}. \quad (2)$$

The $l$, $w$, and $t$ are sample length, width, and thickness, respectively, and the direction of $H_0$ is parallel to the $t$ direction. We prepared a $CoZr_2H_x$ sample with $l$ = 1.22 mm, $w$ = 1.85 mm, and $t$ = 0.49 mm, yielding $N$ = 0.67. Similarly, we prepared a $CoZr_2$ sample with $l$ = 1.43 mm, $w$ = 1.49 mm, and $t$ = 0.32 mm, yielding $N$ = 0.69.

## 3. RESULTS AND DISCUSSIONS

### 3.1 Powder SXRD patterns and crystal structure.

The Rietveld refinement results of powder SXRD patterns for $CoZr_2$ and $CoZr_2H_{3.49}$ at 300 K are displayed in Figure 1a, and the TDS measurement result is shown in Figure 1b. The hydrogen was exhausted mainly as $H_2$ gas, and the rest was detected as H and $H_2O$ gases. The appearance of the two peaks in the TDS spectrum may be caused by differences in coordination numbers and bonding strength between H1/H2 and Co/Zr. The obtained powder SXRD pattern of $CoZr_2$ was successfully indexed using the $CuAl_2$-type tetragonal crystal structure of $I4/mcm$ (no. 140). The powder SXRD pattern of hydrogen-inserted $CoZr_2H_{3.49}$ could also be indexed using a tetragonal crystal structure of $P4/ncc$ (no. 130), which has, however, a lower symmetry crystal space group than $I4/mcm$, as reported in various previous studies.[18,20,24,25] Rietveld refinement results of $CoZr_2H_{3.49}$ and $CoZr_2$ at 300 K, 200 K, and 100 K are shown in Figures S1 and S2. The $CoZr_2$ sample contained $CoZr_3$ (~6% in a mass fraction) and CoZr (~3% in a mass fraction) phases as impurities. The $P4/ncc$ space group is one of the maximal klassengleiche subgroups (k-subgroups), which is obtained by removing a certain translation operation. By inserting hydrogen atoms in $CoZr_2$, $I$-centering is lost, and mirror planes are replaced by $n$- and $c$-glide symmetries. A similar lowering of crystal structural symmetry caused by hydrogen insertion can also be observed in other transition-metal zirconides of $TrZr_2$, such as $FeZr_2$[21] and $NiZr_2$[29]. Refined lattice constants $a$ and $c$ of $CoZr_2H_{3.49}$ at 300 K are $a$ = 6.9331(2) Å and $c$ = 5.6491(3) Å, respectively, and the $c/a$ ratio subsequently calculated as 0.81479(5). For the $CoZr_2$ at 300 K, the $c/a$ ratio is 0.86468(8) with $a$ = 6.3907(5) Å and $c$ = 5.5259(3) Å. Decreasing the $c/a$ ratio in $CoZr_2H_{3.49}$ indicates that an anisotropy of the tetragonal lattice increases by incorporating hydrogen atoms. Indeed, such a tendency has been reported across various hydrogen concentrations.[24] Refined crystal structures of $CoZr_2$ and $CoZr_2H_{3.49}$ are depicted in Figures 1c and 1d, respectively. In the crystal structure of $CoZr_2$, the Co and Zr atoms occupy Wyckoff position 4a and 8h, respectively. The Co layer and Zr layer stack on top of each other along the $c$-axis, and the stacked Co layers form a one-dimensional Co–Co chain along the $c$-axis. Similarly, in $CoZr_2H_{3.49}$, Co layers occupying Wyckoff position 4c and Zr layers occupying Wyckoff position 8f stack on top of each other, with hydrogen atoms incorporated between layers, occupying Wyckoff position 4b and 16g. Therefore, the maximum hydrogen concentration determined by the site occupancy must be $x$ = 5. Co layers build Co–Co chains in the same way as $CoZr_2$. The Co–Co chain length can be calculated by $c/2$ because of the crystal structural symmetry; therefore, the Co–Co chain lengths of $CoZr_2$ and $CoZr_2H_{3.49}$ at 300 K are 2.7630(1) Å and 2.8245(1) Å, respectively. A longer Co-Co chain in CoZr$_2$H$_{3.49}$ leads to a weaker interaction between Co3$dz^2$–Co3$dz^2$ orbitals. The refined atomic coordinates at 300 K are summarized in Tables 1 and 2. For the Rietveld analysis of the powder SXRD pattern of CoZr$_2$H$_{3.49}$, we assume that occupancies of hydrogen atoms are 1, and we apply hydrogen atom coordinates obtained by powder neutron diffraction[20] to our analysis because of the difficulty in accurately deciding hydrogen occupancy and coordinates by synchrotron X-ray.

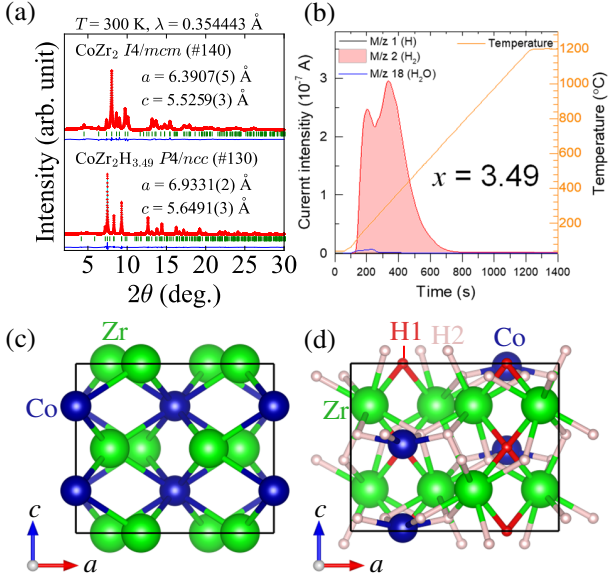

Figure 1. (a) Rietveld refined powder SXRD patterns of CoZr$_2$ and CoZr$_2$H$_{3.49}$. (b) TDS measurement result for CoZr$_2$H$_{3.49}$. (c, d) Depicted crystal structures of CoZr$_2$ and CoZr$_2$H$_{3.49}$ (origin choice 2).

Table 1. Refined atomic coordinates of CoZr$_2$ ($I4/mcm$) at 300 K.

| Label | Atom | x | y | z | Occupancy (fixed) |
|---|---|---|---|---|---|
| Co | Co (4a) | 0 | 0 | 1/4 | 1 |
| Zr | Zr (8h) | 0.17122(7) | 0.67122(7) | 0 | 1 |

Table 2. Refined atomic coordinates of CoZr$_2$H$_{3.49}$ with origin choice 2 ($P4/ncc$) at 300 K. H2 coordination was fixed to the refinement results of powder neutron diffraction data in Ref. 20. Hydrogen occupancies were fixed to 1.

| Label | Atom | x | y | z | Occupancy (fixed) |
|---|---|---|---|---|---|
| Co | Co (4c) | 1/4 | 1/4 | 0.0181(8) | 1 |
| Zr | Zr (8f) | 0.4103(1) | 0.5897(1) | 1/4 | 1 |
| H1 | H (4b) | 3/4 | 1/4 | 0 | 1 |
| H2 | H (16g) | 0.0329 (fixed) | 0.1645 (fixed) | 0.0761 (fixed) | 1 |

### 3.2 Superconductivity and weak-itinerant ferromagnetism of CoZr$_2$H$_x$ ($x$ = 0 and 3.49).

Next, we have investigated the temperature dependencies of $\rho(T)$ and $\chi(T) = M(T)/H$ for CoZr$_2$ and CoZr$_2$H$_{3.49}$. The $\rho(T)$ and $\chi(T)$ curves are shown in Figures 2a–c. The $\chi(T)$ curves were measured on the field cooling (FC) protocol and the zero field cooling (ZFC) protocol. From the $\rho(T)$ of CoZr$_2$, metallic behavior and a superconducting transition at $T_{SC}$ = 6 K were observed. The superconducting transition was also observed in the $4\pi\chi(T)$ curve as shown in Figure 2c. In contrast to the CoZr$_2$ case, CoZr$_2$H$_{3.49}$ exhibited ferromagnetism instead of superconductivity. The $\chi(T)$ curve exhibits paramagnetic behavior at high temperatures and a rapid increment with decreasing temperature, indicating a ferromagnetic transition, as shown in Figure 2b. We observed steep changes in slope for the $\rho(T)$ and $\chi(T)$ curves of CoZr$_2$H$_{3.49}$ near the $T_C$. Determination of the $T_C$ will be discussed later. Changes of temperature dependence of the spin fluctuation part of $\rho(T)$ in the vicinity of critical ferromagnetic transition may be the origin of the kink of the $\rho(T)$ curve of CoZr$_2$H$_{3.49}$ near $T_C$.[30–32] The ferromagnetic transition of CoZr$_2$H$_{3.49}$ can be understood by drastic changes in the electronic density of states (DOS) near the Fermi energy ($E_F$). According to the density functional theory (DFT) calculation,[25] the most drastic change can be seen in the band structure. For CoZr$_2$H$_5$, there is a sharp projected DOS near $E_F$ with major contribution of the antibonding Co3$dz^2$ orbital,[24,25] while the DOS near $E_F$ for CoZr$_2$ is relatively broad, composed of Co3$d$ and Zr4$d$ orbitals.[33] These Co3$d$ and Zr4$d$ orbitals are most likely relevant to the superconductivity in CoZr$_2$. However, in CoZr$_2$H$_5$, the sharp antibonding Co3$dz^2$ orbital positively promotes the emergence of itinerant (band) ferromagnetism in view of the Stoner criterion.[34] The observed ferromagnetic transition of CoZr$_2$H$_{3.49}$ is consistent with not only the theoretical prediction by DFT calculation[25] but also experimental results on different hydrogen-concentration sample of CoZr$_2$H$_{4.8}$.[24]

To precisely determine the $T_C$ of CoZr$_2$H$_{3.49}$, we measured $M(H)$ curves at several temperatures and transformed them to the Arrott plots,[35] which are $M^2$ plots as a function of $H_{eff}/M$. Figure 3a shows the $M(H)$ curve at 1.8 K measured in between −3 and +3 T in a unit of $\mu_B$/f.u. ($\mu_B$ is the Bohr magneton, and f.u. stands for formula unit). CoZr$_2$H$_{3.49}$ exhibited a rapid change in magnetization with a small magnetic field, and the area of the magnetic hysteresis loop is small, which are characteristics of soft magnetic materials. The magnetization was mostly saturated at around 0.4 $\mu_B$. However, the magnetization gradually increased without saturation at least up to ±3 T, which suggests weak-itinerant ferromagnetism,[36,37] same with the CoZr$_2$H$_{4.8}$ sample.[24] Figure 3b shows Arrott plots measured with several steps in 1.8–150 K. Arrott plots are widely used for the determination of $T_C$ in weak-itinerant ferromagnets[36,38–41] because the plots can overcome some of the problems related to the influence of noise in measurements.[38] Because the Arrott plots method is based on Landau mean-field theory, one would expect linear relationships between $M^2$ and $H_{eff}/M$ near $T_C$, and the straight line would pass the origin at $T = T_C$. The straight lines predicted by the Landau mean-field theory would not be observed frequently; nevertheless, we can estimate a reliable $T_C$ value by linear extrapolation of the curve corresponding to a larger value of the magnetization, where sufficiently high magnetic fields are applied.[38] Good linearity between $M^2$ and $H_{eff}/M$ was observed in 138–140 K, as shown in Figure 3c. The black dashed lines represent linear extrapolation to zero field, fitted between 2 and 3 T. The linear extrapolation line passing through the point closest to the origin was obtained at $T = 139$ K within a range of 1 K, suggesting $T_C \sim 139$ K. In addition to the $T_C$ estimation, we can extract temperature dependencies of spontaneous magnetization $M_S(T)$ and inverse initial magnetic susceptibility $\chi_0^{-1}(T)$ ($\chi_0(T) = \lim_{H\to 0} M(T)/H$) by linear extrapolation on the Arrott plots. The $M_S(T)$ and $\chi_0^{-1}(T)$ values were obtained from the interceptions on $M^2$ and $H_{eff}/M$ axes, respectively, by using the linear extrapolation lines fitted between 2 and 3 T on the Arrott plots. Figure 3d shows the temperature dependencies of $M_S(T)$ and $\chi_0^{-1}(T)$. Below $T_C$, we applied the Kuz'min formula[42,43] to fit the $M_S(T)$ curve:

$$M_S(T) = P_S \left[ 1 - s \left(\frac{T}{T_C}\right)^{\frac{3}{2}} - (1-s)\left(\frac{T}{T_C}\right)^{\frac{5}{2}} \right]^{\beta}, \quad (3)$$

where $P_S$ is a spontaneous magnetization at zero field. The $s$ and $\beta$ are dimensionless parameter and critical exponent, respectively. The parameter $s$ must be within the range of $0 < s < 5/2$ for a valid fit.[42,43] As a result of the fitting to $M_S(T)$ curve using Eq. (3), we obtained $P_S = 0.414(4)$ $\mu_B$, $T_C = 139.2(2)$ K, $s = 0.2(1)$, $\beta = 0.58(2)$. From the $T_C$ estimations mentioned above, we determine $T_C = 139$ K for CoZr$_2$H$_{3.49}$. On the other hand, for the above $T_C$, $\chi_0^{-1}(T)$ exhibited linear behavior against temperature. It implies that $\chi_0^{-1}(T)$ obeys the Curie–Weiss behavior, and it is consistent with the self-consistently renormalized (SCR) spin fluctuation theory.[44] Therefore, we use the following general Curie–Weiss law equation for fitting $\chi_0^{-1}(T)$:

$$\chi_0(T) = \frac{C}{T - \theta_{CW}} + \chi_C. \quad (4)$$

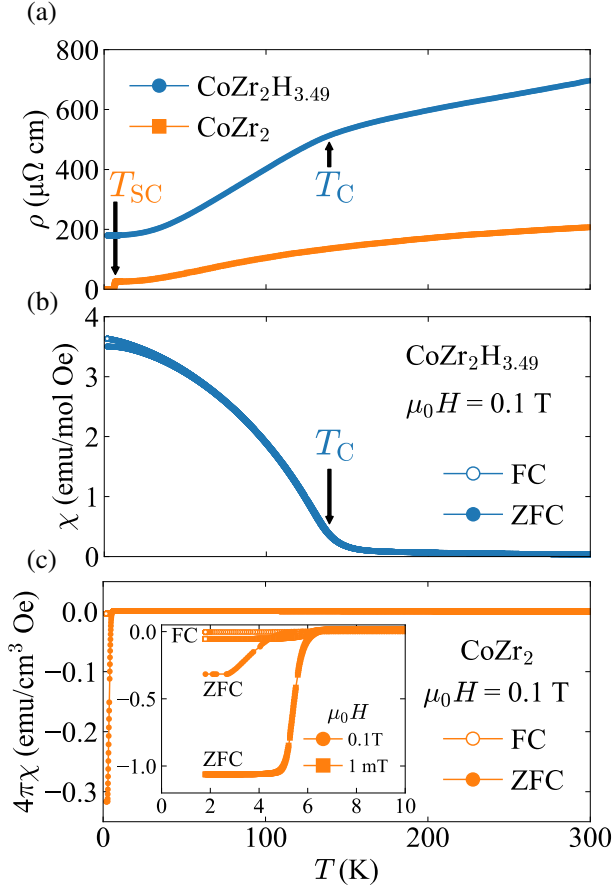

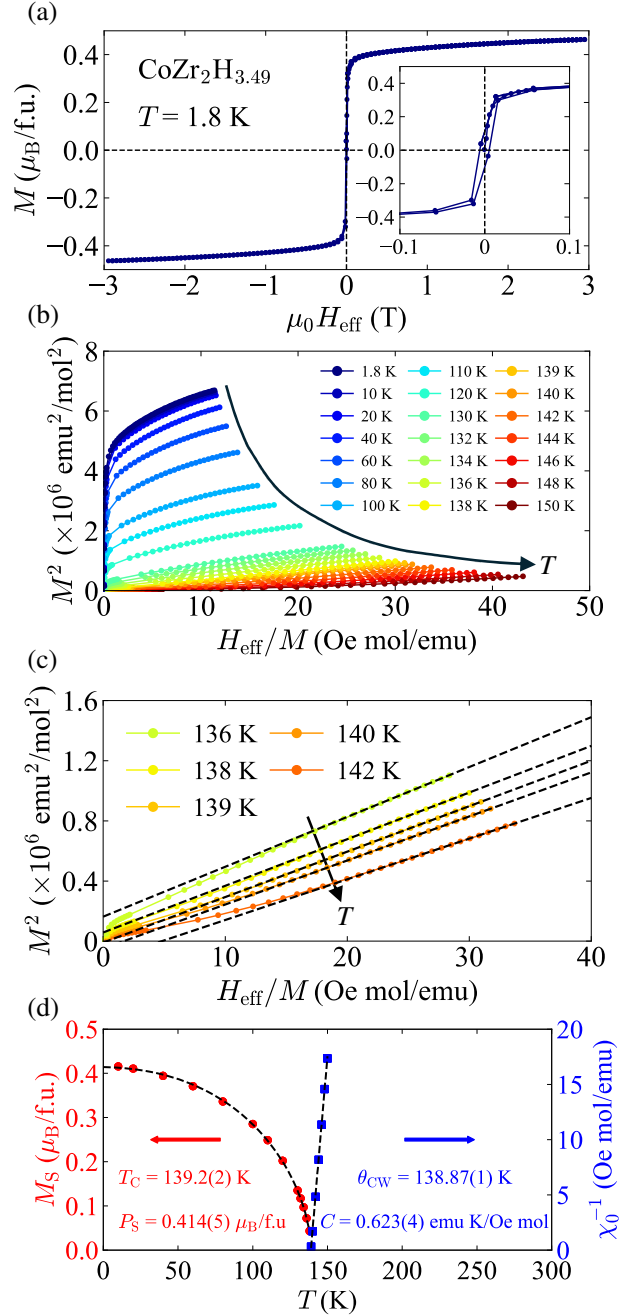

Figure 2. (a) Temperature dependencies of $\rho(T)$ of CoZr$_2$ and CoZr$_2$H$_{3.49}$ under zero field. (b) Temperature dependence of $\chi(T)$ of CoZr$_2$H$_{3.49}$ measured in FC and ZFC protocols under 0.1 T. (c) Temperature dependence of $4\pi\chi(T)$ of CoZr$_2$ measured in FC and ZFC protocols under 0.1 T. The inset shows $4\pi\chi(T)$ curves measured on FC and ZFC protocols near $T_{SC}$ under 1 mT and 0.1 T.

Figure 3. (a) Magnetic field dependence of magnetization for CoZr$_2$H$_{3.49}$ at 1.8 K. The inset figure shows an enlarged view of a magnetic hysteresis loop. (b, c) Arrott plots measured with several steps in 1.8–150 K, and in the vicinity of $T_C$, where 136–142 K. (d) Temperature dependencies of $M_S(T)$ and $\chi_0^{-1}(T)$ obtained from each intercept of Arrott plots on $M^2$ and $H_{eff}/M$ axes, respectively.

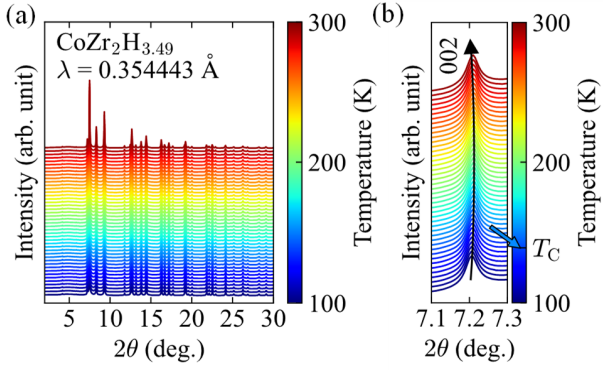

Figure 4. (a) Powder SXRD patterns of CoZr$_2$H$_{3.49}$ measured at several temperatures from 300 K to 100 K. (b) Enlarged view between 7.1° and 7.3°, where only the 002 peak appears. The black solid arrow is a guide to eye for following the position of the 002 peak.

The $C$, $\theta_{CW}$, and $\chi_C$ are the Curie constant, Curie–Weiss temperature, and temperature-independent component, respectively, and we obtained $C = 0.623(4)$ emu K/Oe mol, $\theta_{CW} = 138.87(1)$ K, $\chi_C = 0.0010(4)$ emu/Oe mol. From the estimated $C$ value, we can subsequently calculate the effective magnetic moment as $P_{eff} = 2.232(7)$ $\mu_B$. The number of unpaired electron $P_C$ can be calculated with

$$P_C = \sqrt{1 + P_{eff}^2} - 1, \quad (5)$$

yielding $P_C = 1.446(6)$ $\mu_B$. Finally, we can calculate the Rhodes–Wohlfarth ratio as $P_C/P_S = 3.49(4)$. The Rhodes–Wohlfarth ratio clearly exceeds 1, which indicates CoZr$_2$H$_{3.49}$ exhibits weak-itinerant ferromagnetism. Both CoZr$_2$H$_{4.8}$, reported in Ref. 24, and CoZr$_2$H$_{3.49}$, discussed in this work, exhibit similar magnetic properties; nevertheless, a difference in the $T_C$ is observed. CoZr$_2$H$_{3.49}$ with lower hydrogen concentration exhibits slightly higher $T_C$. In the itinerant ferromagnetic system, $T_C$ is mainly controlled by the width of the conduction band and itinerant charge density, in other words, the Fermi level.[45] The bandwidth consisting of the Co$3dz^2$ orbitals would depend on the Co–Co chain length.[46] Moreover, the charge density can be controlled by hydrogen concentration. We can expect that increasing hydrogen in CoZr$_2$H$_x$ acts as hole doping because of the large electronegativity of hydrogen. According to the DFT calculation in Ref. 25, there is a large charge transfer from Co and Zr towards H1 and H2, which means hydrogen atoms exist with negative charge. Therefore, we can assume that inserting hydrogen into CoZr$_2$H$_x$ plays a role in hole-doping.[24,25] As further important future works, exploring a competition between superconductivity and weak-itinerant ferromagnetism by controlling hydrogen concentration remains.

### 3.3 Uniaxial negative thermal expansion in the weak-itinerant-ferromagnetic phase

The ATE is one of the exotic features of $Tr$Zr$_2$ superconductors, as mentioned in the introduction part. The ATE is primarily attributed to unique atomic motions, which are driven by high-frequency optical phonon branches.[14,47] The striking feature of ATE in CoZr$_2$ superconductor is that it exhibits the uniaxial NTE in a wide temperature range, such as 50–572 K. Thermal expansion properties of CoZr$_2$H$_x$ have not been clarified, but based on the similarity of crystal structures, as we have discussed above, we can expect that the ATE will also be observed in CoZr$_2$H$_{3.49}$. Therefore, we are going to see a change in the lattice constants as a function of temperature. Figure 4a shows the powder SXRD patterns of CoZr$_2$H$_{3.49}$ measured at several temperatures in a range of 100–300 K, covering the border of the paramagnetic and itinerant-ferromagnetic phases. The powder SXRD patterns were collected with a cooling process. $CoZr_2H_{3.49}$ did not exhibit any structural transformation in the examined temperature range. To track changes in the position of 002 peak, in other words, changes in lattice constant $c$, an enlarged view between 7.1° and 7.3° is shown in Figure 4b. Starting from 100 K, it shifted toward the higher angle side, and then it oppositely shifted toward the lower angle side above the $T_C$. This suggests that the temperature dependence of the lattice constant $c$ should differ between the weak-itinerant-ferromagnetic and paramagnetic phases.

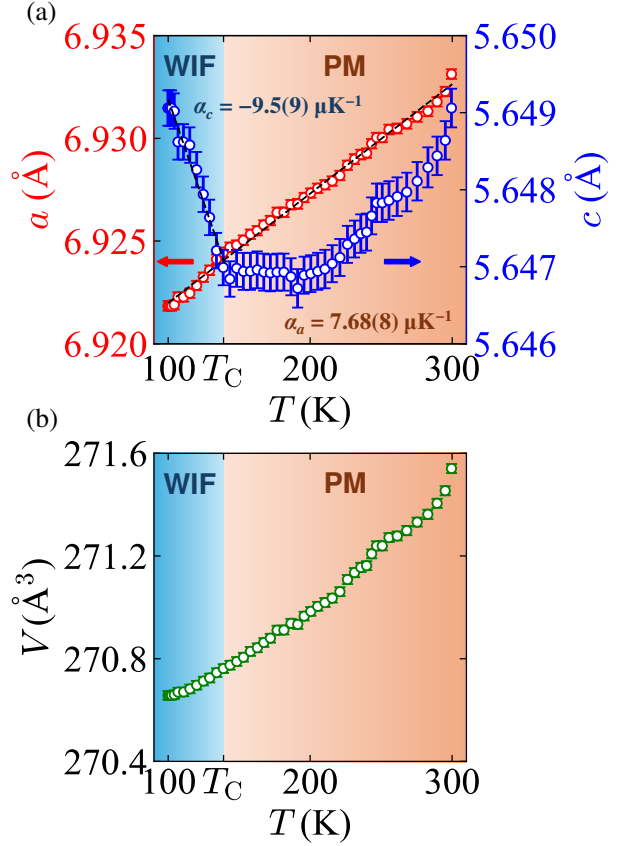

Figure 5. (a) Temperature dependencies of lattice constants $a$ and $c$ for CoZr$_2$H$_{3.49}$ extracted from Rietveld refinements of powder SXRD patterns. WIF and PM stand for weak-itinerant-ferromagnetic and paramagnetic phases, respectively. (b) Temperature dependence of unit cell volume $V = a^2c$.

To clarify ATE, we analyzed each powder SXRD pattern measured at several temperatures using the Rietveld method. We plotted lattice constants $a$ and $c$ as a function of temperature, as shown in Figure 5a. From the analysis, we revealed that $CoZr_2H_{3.49}$ exhibits uniaxial NTE behavior only along the $c$-axis in the weak-itinerant-ferromagnetic phase, while the $a$-axis exhibits usual PTE behavior regardless of the magnetic phase transition. The linear coefficients of thermal expansion $\alpha_l = (dl/dT)/l$ ($l = a$ or $c$) were estimated to be $\alpha_c = -9.5(9)$ μK$^{-1}$ (between 100 K and $T_C$, in the weak-itinerant-ferromagnetic phase) and $\alpha_a = 7.68(8)$ μK$^{-1}$ (between 100 K and 300 K). Although the $c$-axis exhibits NTE, the unit cell volume $V = a^2c$ exhibits PTE as shown in Figure 5b. The studies of the relationship between magnetism and NTE have a long history,[48] which can get back to the discovery of Invar alloys in 1897.[49] Since the discovery of Invar alloys, many magnetic compounds exhibiting NTE have been found,[50-53] and we can classify the origin into six categories: 1. Order-to-disorder transition, 2. Changing local moments, 3. Metamagnetic transition, 4. Magnetic phases coexistence, 5. Short-range magnetic order, and 6. Structural phase transition.[48] The observed ATE in $CoZr_2H_{3.49}$ would belong to the "1. Order-to-disorder transition" because of the transition to the weak-itinerant-ferromagnetic phase below $T_C$. As a consequence of the analysis, we observed the ATE in $CoZr_2H_{3.49}$, similarly to $CoZr_2$. That is, however, there is a clear difference in the ATE feature between them; namely, $CoZr_2H_{3.49}$ exhibits ATE within the weak-itinerant-ferromagnetic phase, while $CoZr_2$ exhibits it in a wide temperature range. We confirmed monotonic NTE behavior along the $c$-axis for $CoZr_2$ (Figure S3). The difference implies that the ATE of $CoZr_2H_{3.49}$ is not driven by phonons, but instead by electrons. Moreover, the fact that only the $c$-axis exhibits NTE indicates that the one-dimensional Co–Co chain bonding running along the $c$-axis plays a key role in the emergence of the ATE.

Herein, we will expand a tentative scenario to understand the ATE of $CoZr_2H_{3.49}$ in the weak-itinerant-ferromagnetic phase. A schematic image of the scenario is displayed in Figure 6. As mentioned above, the weak-itinerant ferromagnetism of $CoZr_2H_{3.49}$ is attributed to the sharp projected DOS near $E_F$ consisting of the antibonding Co$3d_{z^2}$ orbitals. The $d$ bandwidth $W$ is known to be expressed as $W \propto R^{-5}$, where $R$ is the interatomic distance.[46] Therefore, expanding the Co–Co chain length can assist the evolution of weak-itinerant ferromagnetism. This scenario, which is based on the band theory picture, can explain the ATE behavior observed in the weak-itinerant-ferromagnetic phase. The band theory picture has been adopted to explain thermal expansion properties,[8] especially in Invar alloys such as Fe–36Ni[49] and Fe$_3$Pt.[54] However, many approaches have been suggested to explain the Invar effect, which considers spin fluctuations and/or local moment characteristics.[55,56] We may need further unified theories to completely understand the physics of ATE. From the point of view of achieving a giant negative thermal expansion, hybridization of individual mechanisms causing negative thermal expansion would be a key strategy.[8] The anomalous thermal expansion of $CoZr_2$ is caused by high-frequency optical phonons with a flexible framework, likewise Zr/HfW$_2$O$_8$.[7] As a counterpart of that, the origin of the anomalous thermal expansion of $CoZr_2H_{3.49}$ has been revealed as a transition to a weak-itinerant-ferromagnetic phase below $T_C$. We can expect the hybridization of the flexible network and magnetism by adjusting the actual hydrogen concentration near the critical point, where crystal structural transformation between $I4/mcm$ and $P4/ncc$, and competition between superconductivity and ferromagnetism occurs.

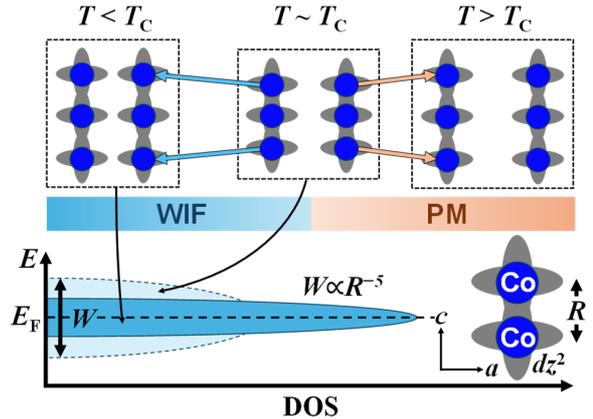

Figure 6. Schematic image of the tentative scenario of the ATE behavior of $CoZr_2H_{3.49}$ in the weak-itinerant-ferromagnetic phase.

## 4. SUMMARY

In this study, we reported that the unique anomalous thermal expansion behavior of $CoZr_2H_{3.49}$. The $CoZr_2H_{3.49}$ exhibited uniaxial NTE behavior along the $c$-axis in the weak-itinerant-ferromagnetic phase. On the other hand, it exhibited usual PTE behavior along the $a$-axis in itinerant ferromagnetic and paramagnetic phases. It is known that $CoZr_2$ also exhibits similar ATE. This is, however, based on a framework structure with high-frequency optical phonons, regardless of the magnetic transition. Therefore, the physical mechanism of the ATE in $CoZr_2H_{3.49}$ reported in this study should be different from the $CoZr_2$ case. The uniaxial NTE along the $c$-axis in the weak-itinerant-ferromagnetic phase can be explained by the basic band theory picture with the Stoner criterion. In detail, expanding the lattice constant $c$ with decreasing temperature below $T_C$ can assist the weak-itinerant ferromagnetism because it promotes a sharper projected DOS near $E_F$ with the major contribution of antibonding Co$3d_{z^2}$ orbital. For further development of these studies, controlling the hydrogen concentration will be important because $CoZr_2H_x$ can serve as an ideal material for studying the competition between superconductivity and ferromagnetism. We can expect hybridization of the individual ATE mechanism of the flexible network and magnetism to achieve giant NTE.

## ASSOCIATED CONTENT

Figure S1: Rietveld refinement results of powder SXRD patterns for $CoZr_2H_{3.49}$ at (a) 300 K, (b) 200 K, and (c) 100 K.

Figure S2: Rietveld refinement results of powder SXRD patterns for CoZr$_2$ at (a) 300 K, (b) 200 K, and (c) 100 K.

Figure S3: (a) Temperature dependencies of lattice constants $a$ and $c$ for CoZr$_2$ extracted from Rietveld refinements of powder SXRD patterns. (b) Temperature dependence of unit cell volume $V = a^2c$.

## AUTHOR INFORMATION

### Corresponding Authors

Yuto Watanabe
Department of Physics, Tokyo Metropolitan University, 1-1 Minami-Osawa, Hachioji, Tokyo 192-0397, Japan
watanabe-yuto@ed.tmu.ac.jp

Yoshikazu Mizuguchi
Department of Physics, Tokyo Metropolitan University, 1-1 Minami-Osawa, Hachioji, Tokyo 192-0397, Japan
mizugu@tmu.ac.jp

### Authors

Kota Suzuki
MDX Research Center for Element Strategy, Institute of Integrated Research, Institute of Science Tokyo, 4259 Nagatsuta, Midori, Yokohama, Kanagawa 226-8501, Japan

Takayoshi Katase
MDX Research Center for Element Strategy, Institute of Integrated Research, Institute of Science Tokyo, 4259 Nagatsuta, Midori, Yokohama, Kanagawa 226-8501, Japan
Materials and Structures Laboratory, Institute of Integrated Research, Institute of Science Tokyo, 4259 Nagatsuta, Midori, Yokohama, Kanagawa 226-8501, Japan

Akira Miura
Faculty of Engineering, Hokkaido University, Kita13, Nishi8, Sapporo, Hokkaido 060-8628, Japan

Aichi Yamashita
Department of Physics, Tokyo Metropolitan University, 1-1 Minami-Osawa, Hachioji, Tokyo 192-0397, Japan

### Notes
The authors declare no competing financial interest.

## ACKNOWLEDGMENT
Y. Watanabe was supported by the Grant-in-Aid for JSPS Fellows (Grant Number JP25KJ1992). Y. Mizuguchi was supported by JSPS KAKENHI (Grant Number JP23KK0088) and TMU research fund for young scientists. T. Katase was supported by Special Award for Science Tokyo Advanced Researchers (STAR) funded by Institute of Science Tokyo.

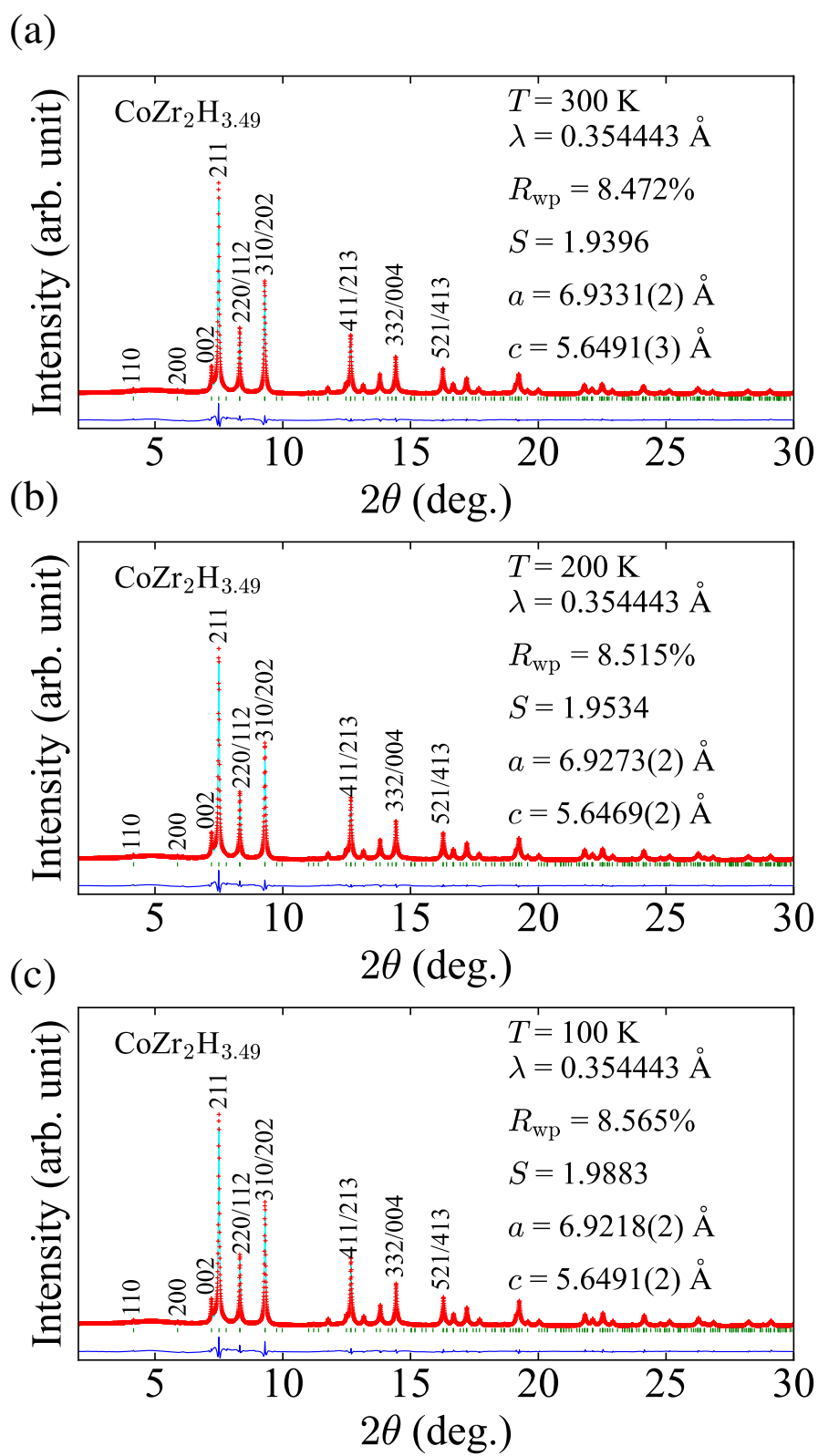

Figure S1. Rietveld refinement results of powder SXRD patterns for CoZr$_2$H$_{3.49}$ at (a) 300 K, (b) 200 K, and (c) 100 K.

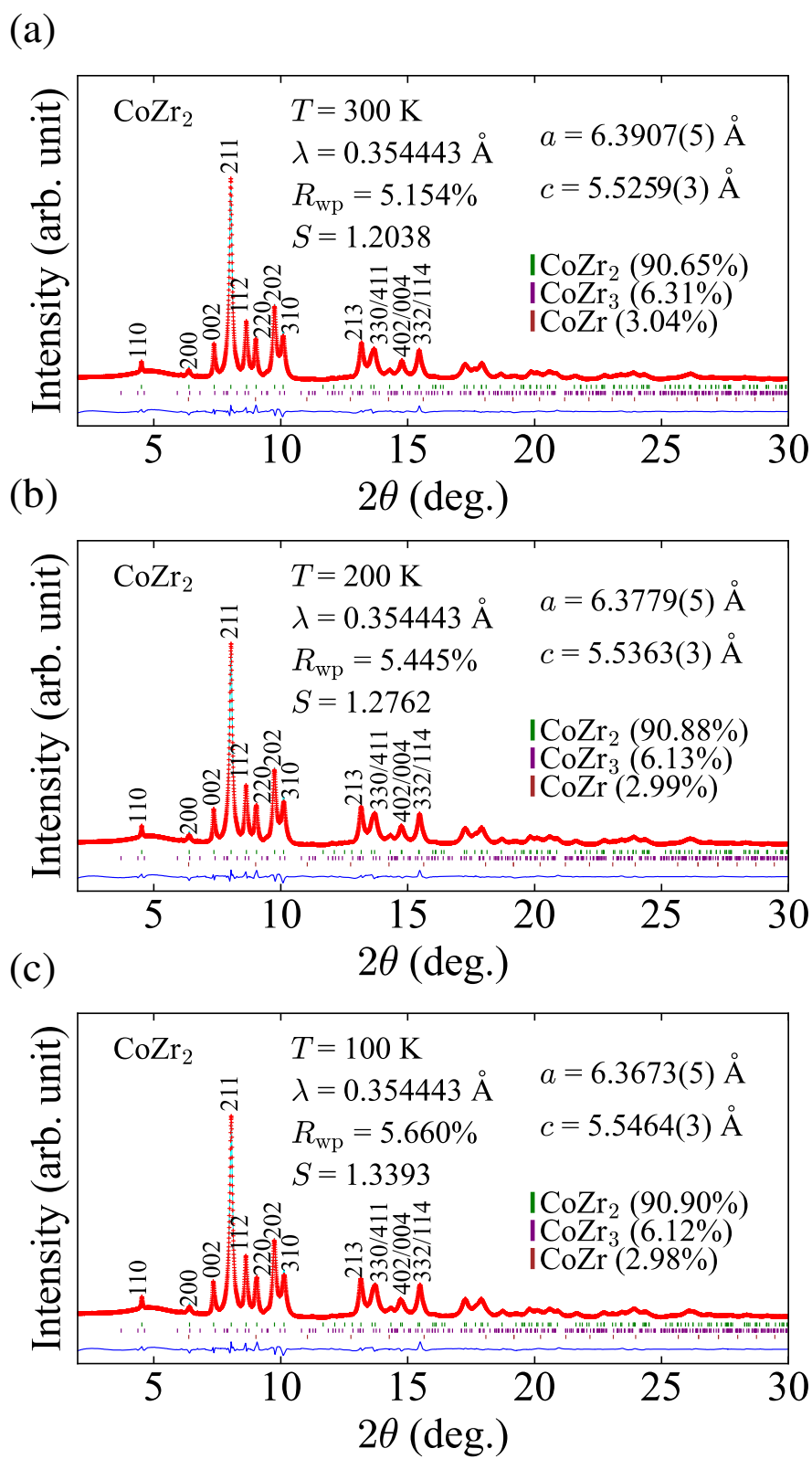

Figure S2. Rietveld refinement results of powder SXRD patterns for CoZr$_2$ at (a) 300 K, (b) 200 K, and (c) 100 K.

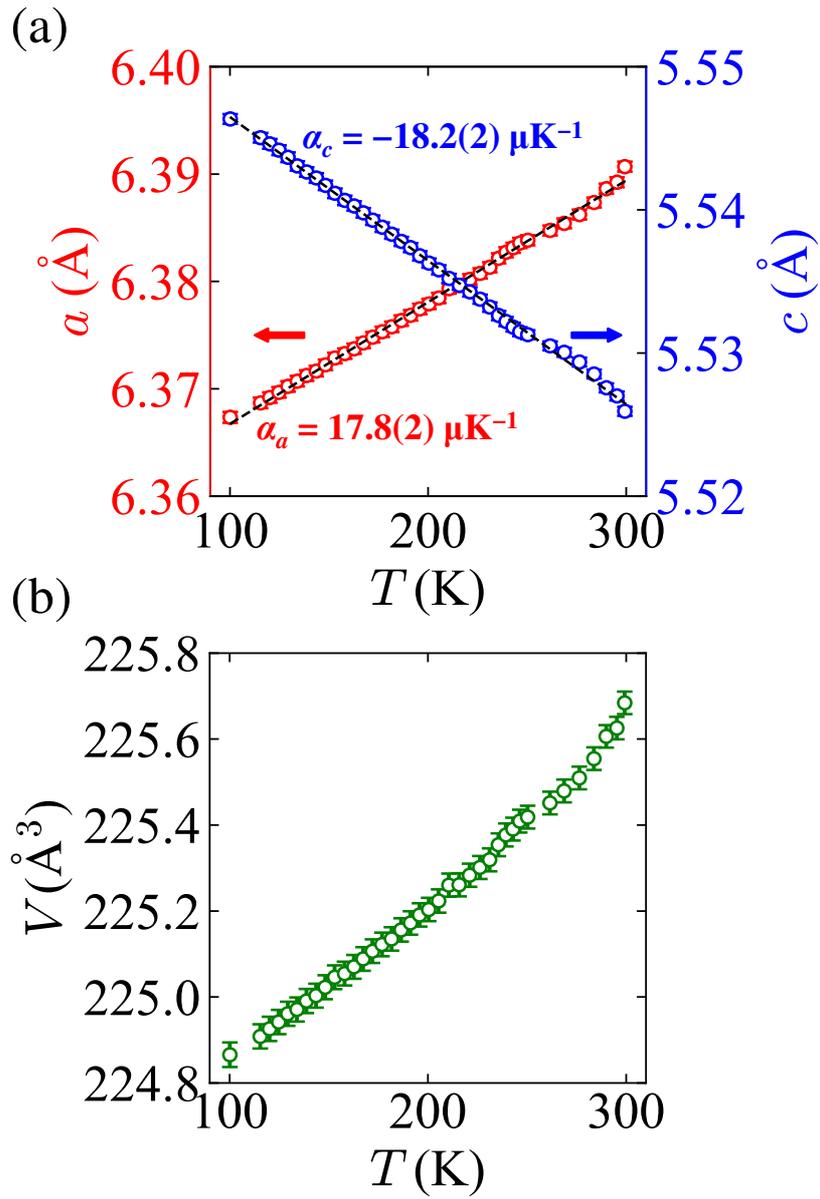

Figure S3. (a) Temperature dependencies of lattice constants $a$ and $c$ for $CoZr_2$ extracted from Rietveld refinements of powder SXRD patterns. (b) Temperature dependence of unit cell volume $V = a^2c$.

12